\documentclass[submission,copyright,creativecommons]{eptcs} 
\providecommand{\event}{WFLP 2016} 

\usepackage{multicol}
\usepackage{amsmath,amsthm,amssymb,cancel}
\usepackage[T1]{fontenc}
\usepackage[utf8]{inputenc}
\usepackage{fancyvrb}
\usepackage{alltt}
\usepackage{textcomp}
\usepackage[english]{babel}
\usepackage{hyperref}
\usepackage{balance}
\usepackage[usenames,dvipsnames,svgnames,table]{xcolor}

\usepackage{listings}
\hypersetup{%
  citecolor=black,
  linkcolor=black,
  urlcolor=black
}

\title{A Framework for Extending microKanren with Constraints}
\author{Jason Hemann \qquad\qquad Daniel P. Friedman
\institute{Indiana University\\Indiana, USA}
\email{\{jhemann,dfried\}@indiana.edu}
}
\def\titlerunning{Framework for Constraints}
\def\authorrunning{J. Hemann, D. P. Friedman}
\begin{document}
\maketitle



\setlength{\pdfpageheight}{\paperheight}
\setlength{\pdfpagewidth}{\paperwidth}








\begin{abstract}
We present a framework for building CLP languages with symbolic
constraints based on microKanren, a domain-specific logic language
shallowly embedded in Racket. We rely on Racket's macro system to
generate a constraint solver and other components of the microKanren
embedding. The framework itself and the constraints' implementations
amounts to just over 100 lines of code. Our framework is both a
teachable implementation for CLP as well as a test-bed and prototyping
tool for symbolic constraint systems.
\end{abstract}





\section{Introduction}\label{introduction}


Constraint logic programming (CLP) is a highly declarative programming
paradigm applicable to a broad class of
problems~\cite{jaffar1994constraint}. Jaffar and Lassez's CLP
scheme~\cite{jaffar87constraint} generalizes the model of logic
programming (LP) to include constraints beyond equations. Constraints
are interpreted as statements regarding some problem domain (e.g., a
term algebra) and a constraint solver evaluates their
satisfiability. The CLP scheme explicitly separates constraint
satisfiability from inference, control, and variable management.

While CLP languages' semantics may respect this separation, their
implementations often closely couple inference and constraint
satisfaction to leverage domain-specific knowledge and improve
performance. Such specialization and tight coupling could force an
aspiring implementer to rewrite a large part of the system to
integrate additional constraints. LP metainterpreters, another common
approach to implementing CLP systems, are unsatisfying in different
ways. The user pays some price for the interpretive overhead, even if
we use techniques designed to mitigate the
cost~\cite{lloyd1991partial}. Moreover, implementing intricate
constraint solvers in a logic language can be unpleasant, and
performance concerns may pressure implementers to sacrifice purity.
The complexity in existing Kanren implementations that support various
symbolic constraints~\cite{comon1994constraints} is also costly in
other ways.

miniKanren is a pure logic language implemented as a purely
functional, shallow embedding in a host language,
e.g. Racket~\cite{flatt2010reference}. microKanren~\cite{hemann2016small}
is an approach to clarifying miniKanren's complexities. It separates
the core implementation from the surface syntax, and is just over 50
lines of code in length. Such separation allows functional programmers
in call-by-value languages to implement the core logic programming
features without the syntactic sugar. There are small syntactic
differences between the miniKanren and microKanren languages and their
various implementations. We elide these details and describe them
collectively as ``Kanrens'' unless otherwise relevant. In all cases a
Kanren embedding gives functional programmers access to logic
programming in languages lacking native support.

The original microKanren is equipped only with a syntactic equality
constraint, but many of the most interesting typical uses of Kanrens
require symbolic constraints~\cite{comon1994constraints} beyond
equality, including disequality constraints, subterm constraints, and
a variety of sort constraints. Implementations with these additions
have ballooned to upwards of a thousand lines of code, and seem
somewhat baroque compared to the 50-line microKanren. Some of this
added heft comes from constraint solving; the remainder involves
transforming constraint sets to solved form and answer projection. In
such versions, implementing each new kind of constraint in the host
language requires plumbing domain knowledge throughout the
implementation on an ad-hoc basis, and complicates constraint solving,
subsumption, reduction to solved form, and answer projection. In short,
it obscures what began as an imminently teachable artifact.

We provide a framework for building shallowly-embedded Racket
implementations of microKanren-like CLP languages with symbolic
constraints. The language designer provides the atomic constraints'
intended interpretations (implemented as Racket predicates) as inputs
to the framework. We then use Racket macros to generate a functional
CLP embedding from the language designer's input. The generated
implementations mirror the separation of inference and solving in the
CLP scheme semantic model. Our macros generate a constraint solver, as
well as other components of the embedding. Our framework's generated
languages return solutions as bags of constraints. We intend to extend
the framework to also reduce constraints to solved, canonical forms,
eliminate redundancies, and project answers with respect to initial
query variables. In the extension we envision, the language designer
describes rewriting-rules for constraint sets, and the framework does
the rest. We envision using our framework as a tool for rapidly
prototyping constraints and CLP languages, in addition to being an
educational artifact for functional programmers.

We describe the Kanren term language, the domain of our symbolic
constraints, in Section~\ref{domain}. We use this framework to
implement common miniKanren constraints in Section~\ref{implementing},
as well as suggestive new ones in Section~\ref{properlisto}. Our
untyped shallow embedding gives the language designer a fair amount of
flexibility, and we describe future directions and alternate design
choices in Section~\ref{conclusion}. This paper is a literate
document; it contains the full implementation of our framework in
Section~\ref{constraints}, and we provide the inference engine in
roughly 30 lines of code in the Appendix. The complete framework and
the implementation of common miniKanren symbolic constraints comprise
just over 100 lines---a marked decrease in line count over similarly
featureful implementations. We also provide our full implementation at
\href{https://github.com/jasonhemann/constraint-microKanren}{\tt
  github.com/jasonhemann/constraint-microKanren} alongside several
syntax extensions implemented as macros on top of the core syntax. To
recapitulate, our contributions include:

\begin{itemize}
\item A macro-based framework for generating pure, functional,
  shallowly-embedded CLP languages in Racket; 
\item The implementations of several symbolic constraints common to Kanren;
\item The implementations of several more suggestive and useful
  symbolic constraints; and 
\item A literate presentation of the complete framework and the
  implementation of our constraints.
\end {itemize}



\section{Background}

We do not expect any miniKanren background or experience of the
reader; a logic programming background is enough. We briefly adumbrate
here the necessary background material. We review some features and
contrast miniKanren's syntax and behavior to that of
Prolog. Interested readers should consult the references, tutorials,
and myriad implementations at \href{www.minikanren.org}{\tt
  miniKanren.org} for further details.

Although Kumar~\cite{kumar2010mechanising} has formally specified
miniKanren's syntax and given several semantics, implementers have
been encumbered by neither, and none of the more widely used
implementations obey these
specifications~\cite{nolen2016core,Byrd:2012:quines}. Instead,
miniKanren is better described as a family of related logic
programming languages, traditionally shallowly embedded in a
declarative host language, most of whose semantics are informally
specified by direct appeal to their host languages' features. Kanrens
inherit much of the syntax and structure of their
hosts. Implementations' concrete syntax varies from host to host
because of the shallow embedding. Different miniKanren implementations
often provide different extensions and operators, as happens with
Prologs. The traditional Prolog definition of a na\"{i}ve reverse
(\verb|nrev|) is syntactically analogous to the miniKanren version
defined using pattern-matching syntax~\cite{mkmacro}.




{\small
\begin{Verbatim}[commandchars=\\\{\},codes={\catcode`$=3\catcode`^=7\catcode`_=8},formatcom={\lstset{fancyvrb=false}}]
                                                 (defmatche (nrev l$\mathtt{_{1}}$ l$\mathtt{_2}$) 
nrev([],[]).                                       ((() ()))
nrev([H|T],L2) :- nrev(T,R),                       (((,h . ,t) ,l$\mathtt{_2}$) 
                  append(R,[H],L2).                 (fresh (r) 
                                                      (nrev t r) 
                                                      (append r `(,h) l$\mathtt{_2}$))))
\end{Verbatim}
}

Rather than using case to distinguish constants and variables, in our
embedding symbols in quoted (\verb|'|) data are constants and are
otherwise considered variables. In pattern matches, we precede
variables by an unquote (\verb|,|), and terms are otherwise considered
constants. Pairs are destructured as
\verb|(,|$\alpha$\verb| . ,|$\beta$\verb|)| rather than
\verb|[|$\alpha$\verb+|+$\beta$\verb|]|. Predicates are defined at
once as a collection of clauses, and we do not require the name of the
predicate in every clause. Unlike in Prolog, the programmer must
explicitly introduce auxiliary variables using the \verb|fresh|
operator.

Below is the translation of the \verb|nrev| into microKanren. The
operators \verb|conj| and \verb|disj| provide binary conjunction and
disjunction; \verb|call/fresh| introduces scope, relying on Racket's
\(\mathtt{\lambda}\) for lexical binding. The operator \verb|==| is microKanren's
first-order syntactic equality constraint. Finally, the operator
\verb|define-relation| defines user predicates and plays a part in the
interleaving. Syntactically, the language resembles Spivey and Seres's
Haskell embedding of Prolog~\cite{spivey1999embedding}. Modulo
differences in syntax, both essentially use \emph{completed
  predicates}, \`{a} la Clark~\cite{clark1978negation}. We can layer
the miniKanren syntactic sugar over this somewhat verbose core syntax
with about 45 lines of Racket macros. This approach is now the most
common way to implement miniKanren.

{\small
\begin{Verbatim}[commandchars=\\\{\},codes={\catcode`$=3\catcode`^=7\catcode`_=8},formatcom={\lstset{fancyvrb=false}}]
(define-relation (nrev l$\mathtt{_1}$ l$\mathtt{_2}$)
  (disj (conj (== l$\mathtt{_1}$ '())
              (== l$\mathtt{_2}$ '()))
        (call/fresh 
          (\(\mathtt{\lambda}\) (h)
            (call/fresh 
              ($\mathtt{\lambda}$ (t)
                (conj (== `(,h . ,t) l$\mathtt{_1}$)
                      (call/fresh 
                        ($\mathtt{\lambda}$ (r)
                          (conj (nrev t r)
                                (append r `(,h) l$\mathtt{_2}$)))))))))))
\end{Verbatim}
}

The \verb|run| operator is the way to evaluate a miniKanren
expression. It takes a number (here \verb|1|), a variable name (here
\verb|q|), and a query to execute. The \verb|run| operator returns a
list of at most that many answers to the query, simplified and answer
projected~\cite{jaffar1993projecting} with respect to the query
variable. Below we demonstrate \verb|nrev| queries in Prolog and
miniKanren.

{\small
\begin{Verbatim}[commandchars=\\\{\},codes={\catcode`$=3\catcode`^=7\catcode`_=8},formatcom={\lstset{fancyvrb=false}}]
?- findnsols(1,L2,nrev([a,b,c],L2),Q).           > (run 1 (q) (nrev '(a b c) q)) 
Q = [[c,b,a]] ?                                  '((c b a))
\end{Verbatim}
}

Prologs default to depth-first search, whereas Kanrens rely on an
unguided, interleaving depth-first search, based on Kiselyov et al.'s
Logic monad~\cite{kiselyov2005backtracking}, that is both complete and
more useful in practice than are breadth-first or iterative deepening
depth-first search. Other than Kanrens, this style of search is not
currently available in any logic language of which we are aware. This
complete, more efficient search makes a purely relational approach to
programming more practical, as we can query pure, all-modes relations
more effectively. This in turn reduces the necessity of non-logical
operators. 

For instance, Kanrens have no equivalent to Prolog's \verb|is|
operator. Instead, the Kanren arithmetic system is built from
all-moded relations built up from bit-adders, without any use of
non-logical operators~\cite{kiselyov2008pure}. The \verb|add| used in
the following query is a 3-place addition relation and \verb|log| is a
four-place logarithm relation. miniKanren prints the answers as
little-endian binary numbers. In the first, we return two answers, and
indeed $0 + 0 = 0$ and $1 + 1 = 2$. In the second, we return a list of
the only answer \verb|(#f #t #t)|, as $14 = 2^{3} + 6$.

{\small
\begin{Verbatim}[commandchars=\\\{\},codes={\catcode`$=3\catcode`^=7\catcode`_=8},formatcom={\lstset{fancyvrb=false}}]
> (run 2 (a b) (add a a b))
'((() ()) ((#t) (#f #t)))
> (run* (q) (log (#f #t #t #t) (#f #t) (#t #t) q))
'((#f #t #t))
\end{Verbatim}
}

These relations run in all modes, and with our interleaving search
appropriately terminate. This declarative arithmetic is efficient
enough to be useful in practice~\cite{brassel2008declaring}. Kanrens
do not currently support arithmetic over real or floating-point
numbers, and heavily numeric computations are not Kanrens' strong
suit. Instead, this all-moded logic programming technique inspires
non-traditional sorts of symbolic computation. Such Kanren programming
examples include a typechecker that also behaves as a type
inhabiter~\cite{alphatap}, an automated theorem prover that doubles as
a proof assistant~\cite{alphamk}, and a programming-language
interpreter that also serves as a quine
generator~\cite{Byrd:2012:quines}. Many of these miniKanren examples
are available in the browser at
\href{http://tca.github.io/veneer/examples/editor.html}{\tt
  tca.github.io/veneer/examples/editor}. Consider as a representative
example \verb|eval|, a relational interpreter for a Racket-like
language. The predicate holds between an expression \verb|e|, an
environment {\tt $\rho$}, and a value \verb|v| when the value of
\verb|e| in {\tt $\rho$} is \verb|v|. To search for a quine, we query
\verb|eval| for an expression that evaluates to its listings (source
code). The result is a valid Racket quine.

{\small
\begin{Verbatim}[commandchars=\\\{\},codes={\catcode`$=3\catcode`^=7\catcode`_=8},formatcom={\lstset{fancyvrb=false}}]
> (run 1 (q) (eval q '() q)
'(((($\mathtt{\lambda}$ (\_.0) (list \_.0 (list 'quote \_.0))) '($\mathtt{\lambda}$ (\_.0) (list \_.0 (list 'quote \_.0))))
   (=/= ((\_.0 closure)))
   (sym \_.0)))
\end{Verbatim}
}

miniKanren prints fresh variables as \verb|_.|$n$ in projected
answers. The above answer is subject to several constraints:
\verb|_.0| must be a symbol other than \verb|closure|. Historically,
miniKanren implementers add new kinds of constraints in response
to challenges that arise in the course of problem solving.

\section{Kanren terms algebra}\label{domain}

The Kanren term language contains symbols, Booleans, the empty list
(\verb|()|), logic variables, and \verb|cons| pairs of the above. In
the interest of simplicity we reserve non-negative integers as logic
variables. Constraints are interpreted directly in the term
structure. Our syntactic equality constraints, built with \verb|==|,
are equations in free \emph{F}-algebra over our logic variables. The
additional common Kanren atomic constraint operators are binary term
disequality, written \verb|=/=|, binary subterm discontainment,
written \verb|absento|, and the unary sort constraints \verb|symbolo|,
and \verb|not-pairo|\footnote{Implementations often also include
  numeric constants and a sort constraint {\tt numbero}. We omit them
  here; their implementation follows naturally.}. These last two
declare the constrained term a symbol or a non-pair respectively. We
term ``constraints'' the finite conjunction of these atomic
constraints.

The meanings of these primitive relation symbols beyond \verb|==| are
given by the language designer as a collection of host-language
predicates. We first decide the satisfiability of the conjunction of
the \verb|==| constraints, either failing or producing a
substitution. Relying on the independence of negative constraints, we
then use the provided predicates to determine the satisfiability of
the remaining constraints modulo that substitution. We check all
atomic constraints of the same kind at once, checking each kind in
turn.


In addition to their usual benefits, our constraints also allow us to
compress what would be multiple answers (potentially infinitely many)
into single finite representations. Consider for instance, the
\verb|absento| constraint. An \verb|absento| constraint holds between
two terms \verb|x| and \verb|y| when \verb|x| is neither equal to, nor
a subterm of \verb|y|. With just \verb|=/=| constraints, we can in
general only express this relationship in the limit, e.g. an infinite
conjunction of disequalities between the fresh variable \verb|y| and
all possible terms \verb|x| from which it is absent. With the
\verb|absento| constraint we can represent this relationship finitely.









\section{Constraint framework user's requirements}\label{restrictions}

A CLP-language designer is the \emph{user} of our framework. They provide
as inputs to the framework's macros their language's atomic constraint
relation symbols, as Racket symbols, and the conditions that violate
constraints, via predicates to test for invalid sets of
constraints. From these, the framework will generate the microKanren
(and thus miniKanren) constraint operators and a constraint solver
automatically. They are a declarative, functional specification of
what it means to violate these constraints. Constraint-violation
predicates, qua predicates, are by definition total functions. As
such, the solver for the constraint system (\verb|invalid?|, defined
in Section~\ref{constraints}) is also total.




We provide \verb|==|, representing syntactic first-order equality,
with every constraint system, and implement it via unification with
\verb|occurs?| check (see Appendix). We fix this particular equational
theory because of our intended use cases for the generated languages
(e.g. sound type inference in a simply-typed language).

We require the resultant constraint solver to be
\textit{well-behaved}~\cite{jaffar1998semantics}. This means it is
\textit{logical}---that is, it gives the same answer for any
representation of the same constraint information (i.e., regardless of
order, redundancy, etc). It is also \textit{monotonic}---that is, for
any set of constraints, if the solver deems the set invalid, adding
additional constraints cannot produce a valid set. Therefore, when
adding a new constraint-violation predicate, a language designer need
not modify older ones. Such a redesign may, however, clarify these
violations. Presently, these requirements are unchecked.

\enlargethispage*{\baselineskip}

\section{Constraint framework implementation}\label{constraints}
 
In this section we completely implement our framework. We implement
the constraint store as a persistent hash table in the host
language. Each type of atomic constraint operator in the embedded
language is a distinct key/value pair in the hash table. We use the
relation symbol as that relation's key in the store. We define the
initial state (constraint store) \(\mathtt{S_0}\) as an immutable hash table
with \verb|==| and each of the provided constraint identifiers as keys
associated with \verb|()|, Racket empty lists.

{\small
\begin{Verbatim}[commandchars=\\\{\},codes={\catcode`$=3\catcode`^=7\catcode`_=8},formatcom={\lstset{fancyvrb=false}}]
(define S\(\mathtt{_0}\) (make-immutable-hasheqv '((==) (cid) ...)))
\end{Verbatim}
}

Since symbols for distinct relations are unique, each field will have
a distinct key. The hash table is immutable to allow structure sharing
across different extensions of the same store, and we rely on the host
language's garbage collection to free memory.


{\small
\begin{Verbatim}[commandchars=\\\{\},codes={\catcode`$=3\catcode`^=7\catcode`_=8},formatcom={\lstset{fancyvrb=false}}]
> (define == (make-constraint-goal-constructor '==))
  ...
\end{Verbatim}
}

Invoking \verb|make-constraint-goal-constructor| yields the
implementation of each relation in our embedding. We use the
relation's name as its key in the store.
\verb|make-constraint-goal-constructor| takes a field in the store and
returns a function accepting the correct number of term
arguments. Such functions are the definitions of relations in our
embedding.

{\small
\begin{Verbatim}[commandchars=\\\{\},codes={\catcode`$=3\catcode`^=7\catcode`_=8},formatcom={\lstset{fancyvrb=false}}]
(define (((make-constraint-goal-constructor key) . ts) S/c)
  (let ((S (ext-S (car S/c) key ts)))
    (if (invalid? S) '() (list `(,S . ,(cdr S/c))))))
\end{Verbatim}
}

Invoking this relation with terms yields a goal: a function expecting
a store and returning a stream of stores. To evaluate a constraint we
extend the store by adding the newly constrained terms and test the
store's consistency. If the extended constraint store is consistent,
we return a stream of a single store; if not, we return the empty
stream. Once added, constraints are not removed from the store. This
decision means the size of the constraint store and the cost of
checking constraints grows each time we encounter a constraint in the
execution of a program. In Section~\ref{conclusion} we suggest
improvements.

The \verb|ext-S| function takes the store, the key, and a list of
terms. The \verb|ext-S| function adds those terms, as a data
structure, to a list of such structures. By \verb|cons|ing all of the
terms together, \verb|hash-update| creates the data structure.

{\small
\begin{Verbatim}[commandchars=\\\{\},codes={\catcode`$=3\catcode`^=7\catcode`_=8},formatcom={\lstset{fancyvrb=false}}]
(define (ext-S S key ts) (hash-update S key ((curry cons) (apply list* ts))))
\end{Verbatim}
}

We check consistency with \verb|invalid?|. \verb|make-invalid?| builds
the definition of \verb|invalid?|. The language designer provides
\verb|make-invalid?| a list of the relation symbols (Racket
identifiers). The designer also provides a sequence of predicates that
check for constraint violations. Each predicate takes a substitution
and returns true if it detects a violation. The constraint identifiers
are free variables of the predicates; the expansion of
\verb|make-invalid?| will bind them. The result of
\verb|make-invalid?| is a predicate that tests if a store is invalid.

The Racket primitive \verb|define-syntax-rule| builds a macro. This
macro transforms an occurrence of the pattern, an expression beginning
with \verb|make-initial-state| followed by zero or more identifiers
into an instantiation of the macro's template.

{\small
\begin{Verbatim}[commandchars=\\\{\},codes={\catcode`$=3\catcode`^=7\catcode`_=8},formatcom={\lstset{fancyvrb=false}}]
(define-syntax-rule (make-invalid? (cid ...) p ...)
  (\(\mathtt{\lambda}\) (S) (let ((cid (hash-ref S 'cid)) ...)
           (cond ((valid-== (hash-ref S '==)) => (\(\mathtt{\lambda}\) (s) (or (p s) ...)))
                 (else #t)))))
\end{Verbatim}
}

\noindent The first relation we check is \verb|==|. If these
constraints are consistent, the result is a substitution that is the
m.g.u.~of these constraints. Assuming this field is valid, we pass the
resulting substitution as an argument to the constraint-violation
predicates.

Our framework includes the implementation of the relation \verb|==|
and provides \verb|==| in every generated constraint system. The
\verb|==| relation is special because we consider the satisfiability
of other relations modulo this equivalence. For the remaining
relations, we can apply the substitution to the constrained terms and
check for satisfiability in the free term algebra generated by the set
of variables away from the domain of the substitution. The
\verb|valid-==| function below and its associated help functions are
also included with the framework. The \verb|valid-==| function expects
a list of cons pairs of terms to unify with each other. We define
\verb|unify| in the Appendix.

{\small
\begin{Verbatim}[commandchars=\\\{\},codes={\catcode`$=3\catcode`^=7\catcode`_=8},formatcom={\lstset{fancyvrb=false}}]
(define (valid-== ==) 
  (foldr (\(\mathtt{\lambda}\) (pr s) (and s (unify (car pr) (cdr pr) s))) '() ==))
\end{Verbatim}
}

This is the main syntactic form for building constraint systems. We
build the entire constraint system and embedded language with one
invocation of \verb|make-constraint-system|. This new syntactic form
takes the same parameters as does \verb|make-invalid?|. It builds
\verb|invalid?|, the initial store, and all the functions implementing
the relations themselves. The result is a constraint system; together
with microKanren's control infrastructure (see Appendix) this yields a
full implementation of a microKanren-like CLP language. To construct a
microKanren with just equality, the language designer invokes
\verb|make-constraint-system| with an empty list of relation
identifiers and no violation predicates.

{\small
\begin{Verbatim}[commandchars=\\\{\},codes={\catcode`$=3\catcode`^=7\catcode`_=8},formatcom={\lstset{fancyvrb=false}}]
> (make-constraint-system ())
\end{Verbatim}
}

The definition below uses Racket's
\verb|syntax-parse|~\cite{culpepper2010fortifying}, a more
sophisticated macro system. We pattern-match on the syntax argument,
and the hash (\verb|#|) begins the definition of the syntax template.
We use \verb|syntax-local-introduce| to introduce three new
identifiers into lexical scope; the remaining constraint identifiers
are already scoped.

{\small
\begin{Verbatim}[commandchars=\\\{\},codes={\catcode`$=3\catcode`^=7\catcode`_=8},formatcom={\lstset{fancyvrb=false}}]
(define-syntax (make-constraint-system stx)
  (syntax-parse stx
    [(\_ (cid:id ...) p ...)
     (with-syntax ([invalid? (syntax-local-introduce #'invalid?)]
                   [S\(\mathtt{_0}\) (syntax-local-introduce #'S\(\mathtt{_0}\))]
                   [== (syntax-local-introduce #'==)])
       #'(begin (define invalid? (make-invalid? (cid ...) p ...))
                (define S\(\mathtt{_0}\) (make-immutable-hasheqv '((==) (cid) ...)))
                (define == (make-constraint-goal-constructor '==))
                (define cid (make-constraint-goal-constructor 'cid))
                ...))]))
\end{Verbatim}
}

\noindent This macro is the primary driver of our framework. The
preceding code and the half page of code in the Appendix comprise the
entire implementation.

\section{Implementing a constraint system}\label{implementing}

Next, we make further use of our framework. We implement a series of
violation predicates with some associated help functions and use those
predicates to generate a constraint system for common symbolic
constraints. We develop these relations and their predicates one at a
time.

The typical Kanren contains four other relations beyond
\verb|==|. These are \verb|=/=|, \verb|absento|, \verb|symbolo|, and
\verb|not-pairo|. We discuss the predicates required to implement
these relations one at a time.

We first add a predicate to test for a violated \verb|=/=|
constraint. This predicate searches for an instance where, with
respect to the current substitution, two terms under a \verb|=/=|
constraint are already equal. In that case, the \verb|=/=| constraint is
deemed violated.

{\small
\begin{Verbatim}[commandchars=\\\{\},codes={\catcode`$=3\catcode`^=7\catcode`_=8},formatcom={\lstset{fancyvrb=false}}]
> (make-constraint-system (=/= absento symbolo not-pairo)
    (\(\mathtt{\lambda}\) (s) (ormap (\(\mathtt{\lambda}\) (pr) (same-s? (car pr) (cdr pr) s)) =/=))
    ...)
\end{Verbatim}
}

We implement this predicate in terms of a help function
\verb|same-s?|. If the result of unifying two terms in the
substitution is the same as the original substitution, then those
terms were already equal relative to that substitution.

{\small
\begin{Verbatim}[commandchars=\\\{\},codes={\catcode`$=3\catcode`^=7\catcode`_=8},formatcom={\lstset{fancyvrb=false}}]
#| Term \(\mathtt{\times}\) Term \(\mathtt{\times}\) Subst \(\mathtt{\rightarrow}\) Bool |#  
(define (same-s? u v s) (equal? (unify u v s) s))
\end{Verbatim}
}

The next predicate checks for violated \verb|absento| constraints,
using the auxiliary predicate \verb|mem?|. The predicate searches for
an instance where, with respect to the substitution, the first term of
a pair is already equal to (a subterm of) the second term. In that
case, we deem the \verb|absento| constraint violated.

{\small
\begin{Verbatim}[commandchars=\\\{\},codes={\catcode`$=3\catcode`^=7\catcode`_=8},formatcom={\lstset{fancyvrb=false}}]
> (make-constraint-system (=/= absento symbolo not-pairo)
    ...
    (\(\mathtt{\lambda}\) (s) (ormap (\(\mathtt{\lambda}\) (pr) (mem? (car pr) (cdr pr) s)) absento))
    ...)
\end{Verbatim}
}

The predicate \verb|mem?| checks if a term \verb|u| is already
equivalent to any subterm of a term \verb|v| under a substitution
\verb|s|. It makes use of \verb|same-s?| in the check. If the result
of unifying \verb|u| and \verb|v| is the same as the substitution
\verb|s| itself, then the two terms are already equal.

{\small
\begin{Verbatim}[commandchars=\\\{\},codes={\catcode`$=3\catcode`^=7\catcode`_=8},formatcom={\lstset{fancyvrb=false}}]
#| Term \(\mathtt{\times}\) Term \(\mathtt{\times}\) Subst \(\mathtt{\rightarrow}\) Bool |#  
(define (mem? u v s)
  (let ((v (walk v s)))
    (or (same-s? u v s) (and (pair? v) (or (mem? u (car v) s) (mem? u (cdr v) s))))))
\end{Verbatim}
}

We write a third violation predicate to search for a violated
\verb|symbolo| constraint. For each term under a \verb|symbolo|
constraint, we look if that term, relative to the substitution, is
anything but a symbol or a variable. If so, that term violates the
constraint. We define \verb|walk|, a function that performs a deep
lookup of a term in a substitution, in the Appendix.

{\small
\begin{Verbatim}[commandchars=\\\{\},codes={\catcode`$=3\catcode`^=7\catcode`_=8},formatcom={\lstset{fancyvrb=false}}]
> (make-constraint-system (=/= absento symbolo)
    ...
    (\(\mathtt{\lambda}\) (s) (ormap (\(\mathtt{\lambda}\) (y)
                    (let ((t (walk y s)))
                      (not (or (symbol? t) (var? t)))))
             symbolo))
    ...)
\end{Verbatim}
}

The \verb|not-pairo| violation predicate operates similarly. This last
definition completes our implementation of the symbolic constraints
common to Kanrens.

{\small
\begin{Verbatim}[commandchars=\\\{\},codes={\catcode`$=3\catcode`^=7\catcode`_=8},formatcom={\lstset{fancyvrb=false}}]
> (make-constraint-system (=/= absento symbolo not-pairo)
    ...
    (\(\mathtt{\lambda}\) (s) (ormap (\(\mathtt{\lambda}\) (n)
                    (let ((t (walk n s)))
                      (not (or (not (pair? t)) (var? t)))))
             not-pairo)))
\end{Verbatim}
}



We show below the execution of an example microKanren program that
uses all the typical kinds of Kanren constraints. The result of
invoking this program is a stream containing a single store. We see
that all the constraints are present in the constraint store, and we
can read off each constraint. The \verb|#hasheqv(...)| is the printed
representation of the hash table, whose elements are the key/value
pairs. For instance, the \verb|=/=| field,
\verb|(=/= . ((c . 0) (0 . b)))|, contains the pairs \verb|(c . 0)|
and \verb|(0 . b)|. These are the \verb|=/=| constraints that have
been added.

{\small
\begin{Verbatim}[commandchars=\\\{\},codes={\catcode`$=3\catcode`^=7\catcode`_=8},formatcom={\lstset{fancyvrb=false}}]
> (call/initial-state 1
    (call/fresh 
      (\(\mathtt{\lambda}\) (x)
        (conj (== 'a x)
              (conj (=/= x 'b)
                    (conj (absento 'b `(,x))
                          (conj (not-pairo x)
                                (conj (symbolo x)
                                      (=/= 'c x)))))))))
'((#hasheqv((== . ((a . 0))) (=/= . ((c . 0) (0 . b))) (absento . ((b 0)))
            (symbolo . (0)) (not-pairo . (0)))
   . 1))
\end{Verbatim}
}

\section{Adding new constraints}\label{properlisto}

Beyond refactoring existing implementations, our framework also
simplifies describing more complicated symbolic constraints new to
Kanren: \verb|booleano| and \verb|listo|. The first mandates that the
constrained term be a Boolean, and the second a proper list. These
relations have more complex interactions than do the previous ones,
and we need several new predicates to support each of these relations'
implementation.

We add support for these constraints both because of their additional
complexity and also their utility. With them, we can improve the
implementations of relational interpreters, one of the archetypal
miniKanren programming examples. Consider the partially-completed
miniKanren definition of the relational interpreter \verb|eval| below.

{\small
\begin{Verbatim}[commandchars=\\\{\},codes={\catcode`$=3\catcode`^=7\catcode`_=8},formatcom={\lstset{fancyvrb=false}}]
(defmatche (eval e \(\mathtt{\rho}\) v)
  ((,e ,\(\mathtt{\rho}\) ,v) (fresh () (symbolo e) (lookup e \(\mathtt{\rho}\) v)))
  ((,e ,\(\mathtt{\rho}\) ,v) (fresh () (booleano e) (listo \(\mathtt{\rho}\))))
  ...)
\end{Verbatim}
}

If \verb|e| is a variable, \verb|v| is its value in the environment
\(\mathtt{\rho}\). We define \verb|lookup| recursively as a
three-place user predicate. When the variable is found in the
environment, we return its value. In prior implementations of
relational interpreters, the remainder of the environment remains
unconstrained. Without \verb|listo| constraints, the only way to
ensure environments are proper lists requires generating via yet
another user predicate the proper lists of all given lengths. Instead,
we can now express infinitely many answers with a single \verb|listo|
constraint. We have more tightly constrained the implementation of
\verb|lookup|, which results in more precise answers.

{\small
\begin{Verbatim}[commandchars=\\\{\},codes={\catcode`$=3\catcode`^=7\catcode`_=8},formatcom={\lstset{fancyvrb=false}}]
(defmatche (lookup x \(\mathtt{\rho}\) o)
  ((,x ((,x . ,o) . ,d) o) (listo d))
  ((,x ((,aa . ,da) . ,d) o) (fresh () (=/= aa x) (lookup x d o))))
\end{Verbatim}
}

In prior definitions of \verb|eval|~\cite{Byrd:2012:quines}, rather
than using a \verb|booleano| constraint, we equated the term first
with \verb|#t|, and then separately with \verb|#f|. This generates
near-duplicate programs that differ in their placement of \verb|#t|
and \verb|#f|. By instead ``compressing'' the Booleans into one, we
ensure the programs we generate have a more interesting variety.

\subsection{Implementing \tt{booleano}}

Checking \verb|booleano| involves more work than does checking our
earlier sort constraints, since Boolean values are sort limited. The
first predicate holds if we have forbid a term from being either of
the constants \verb|#t| and \verb|#f| while demanding that it be a
Boolean. We also need a predicate to check for a
\verb|booleano|-constrained term that is a non-variable,
non-Boolean. Finally since the \verb|booleano| domain constraint is
incompatible with \verb|symbolo|, the last predicate checks for terms
constrained by both.

{\small
\begin{Verbatim}[commandchars=\\\{\},codes={\catcode`$=3\catcode`^=7\catcode`_=8},formatcom={\lstset{fancyvrb=false}}]
> (make-constraint-system (=/= absento symbolo not-pairo booleano)
    ...
    (let ((not-b (\(\mathtt{\lambda}\) (s) (or (ormap (\(\mathtt{\lambda}\) (pr) (same-s? (car pr) (cdr pr) s)) =/=)
                            (ormap (\(\mathtt{\lambda}\) (pr) (mem? (car pr) (cdr pr) s)) absento)))))
      (\(\mathtt{\lambda}\) (s) (ormap (\(\mathtt{\lambda}\) (b) (let ((s\(\mathtt{_1}\) (unify b #t s)) (s\(\mathtt{_2}\) (unify b #t s)))
                             (and s\(\mathtt{_1}\) s\(\mathtt{_2}\) (not-b s\(\mathtt{_1}\)) (not-b s\(\mathtt{_2}\)))))
               booleano)))
    (\(\mathtt{\lambda}\) (s) (ormap (\(\mathtt{\lambda}\) (b) (let ((b (walk b s)))
                           (not (or (var? b) (boolean? b)))))
             booleano))
    (\(\mathtt{\lambda}\) (s) (ormap (\(\mathtt{\lambda}\) (b) (ormap (\(\mathtt{\lambda}\) (y) (same-s? y b s)) symbolo))
             booleano)))
\end{Verbatim}
}

\noindent The following is an example of its use.

{\small
  \begin{Verbatim}[commandchars=\\\{\},codes={\catcode`$=3\catcode`^=7\catcode`_=8},formatcom={\lstset{fancyvrb=false}}]
> (call/initial-state 1
    (call/fresh 
      (\(\mathtt{\lambda}\) (x)
        (conj (=/= #f x)
              (conj (=/= #t x)
                    (booleano x))))))
'()
\end{Verbatim}
}

\subsection{Implementing \tt{listo}}

Checking \verb|listo| is more complicated still. Consequently some of
the violation predicates are also quite complex. We add four
independent predicates to properly implement \verb|listo|. We briefly
descibe their behavior, and then provide their implementations. 

In the first of these functions, we look for an instance in which the
end of a term labeled a proper list \verb|l| is required to be a
symbol. We use the help function \verb|walk-to-end| in
constraint-violation predicates related to \verb|listo| constraints.
This help function recursively walks the \verb|cdr| of a term \verb|x|
in a substitution \verb|s| and returns the final \verb|cdr| of
\verb|x| relative to \verb|s|.

{\small
\begin{Verbatim}[commandchars=\\\{\},codes={\catcode`$=3\catcode`^=7\catcode`_=8},formatcom={\lstset{fancyvrb=false}}]
#| Term \(\mathtt{\times}\) Subst \(\mathtt{\rightarrow}\) Bool |#  
(define (walk-to-end x s)
  (let ((x (walk x s)))
    (if (pair? x) (walk-to-end (cdr x) s) x)))
\end{Verbatim}
}

The second predicate resembles the first, except it checks for a
Boolean instead. In the third, we check for a proper list that must
have a definite fixed last cdr (the \verb|end|) under the
substitution. This means either \verb|end| already is \verb|()|, or a
\verb|not-pairo| constrains \verb|end|. If, in addition, either
\verb|=/=| or \verb|absento| constraints forbid \verb|end| from being
\verb|()|, then that is a violation.

In the last predicate we require in order to correctly implement
\verb|listo|, \verb|end| can be a proper list of unknown length. An
\verb|absento| constraint forbidding \verb|()| from occurring in a
term containing \verb|end|, however, causes a violation. The
constraint must precisely forbid \verb|()| from occurring in a term
containing \verb|end| to cause the violation. 



{\small
\begin{Verbatim}[commandchars=\\\{\},codes={\catcode`$=3\catcode`^=7\catcode`_=8},formatcom={\lstset{fancyvrb=false}}]
> (make-constraint-system (=/= absento symbolo not-pairo booleano listo)
    ...
    (\(\mathtt{\lambda}\) (s) (ormap (\(\mathtt{\lambda}\) (l) (let ((end (walk-to-end l s)))
                           (ormap (\(\mathtt{\lambda}\) (y) (same-s? y end s)) symbolo)))
             listo))
    (\(\mathtt{\lambda}\) (s) (ormap (\(\mathtt{\lambda}\) (l) (let ((end (walk-to-end l s)))
                           (ormap (\(\mathtt{\lambda}\) (b) (same-s? b end s)) booleano)))
             listo))
    (\(\mathtt{\lambda}\) (s) (ormap (\(\mathtt{\lambda}\) (l) (let ((end (walk-to-end l s)))
                           (let ((\^{s} (unify end '() s)))
                             (and \^{s}
                                  (ormap (\(\mathtt{\lambda}\) (n) (same-s? end n s)) not-pairo)
                                  (or (ormap (\(\mathtt{\lambda}\) (pr) (same-s? (car pr) (cdr pr) \^{s}))
                                        =/=)
                                      (ormap (\(\mathtt{\lambda}\) (pr) (mem? (car pr) (cdr pr) \^{s}))
                                        absento))))))
             listo))
    (\(\mathtt{\lambda}\) (s) (ormap (\(\mathtt{\lambda}\) (l) (let ((end (walk-to-end l s)))
                           (ormap (\(\mathtt{\lambda}\) (pr) (and (null? (walk (car pr) s))
                                               (mem? end (cdr pr) s)))
                             absento)))
             listo))
    ...)
\end{Verbatim}
}


These violation predicates are somewhat involved---of necessity. We
have ensured that constraint violations can each be treated
independently and that they comprise the entirety of the constraint
domain knowledge required. Furthermore, by requiring that our solver
be monotonic and logical, we have ensured that adding new constraints
never requires the language designer to modify existing
predicates. Below is an example of a sample constraint microKanren
programs using these new forms of constraints. We can still provide
miniKanren syntax to the language user with with but a handful of
host-language macros~\cite{hemann2016small}.

{\small
\begin{Verbatim}[commandchars=\\\{\},codes={\catcode`$=3\catcode`^=7\catcode`_=8},formatcom={\lstset{fancyvrb=false}}]
> (call/initial-state 1
    (call/fresh 
      (\(\mathtt{\lambda}\) (x)
        (conj (listo x)
              (conj (not-pairo x)
                    (disj (=/= '() x)
                          (absento x '())))))))
'()
\end{Verbatim}
}


\section{Related work}\label{related}


The modern development of CLP languages begins in the mid 1980s by
groups in Melbourne, Marseilles, and the ECRC. The CLP
scheme~\cite{jaffar87constraint} is an important development from this
era. The CLP scheme separates the inference mechanism from
constraint handling and satisfaction. It subsumes many individual
logic programming extensions and provides a theoretical foundation for
disparate CLP languages.

CLP over the domain of finite trees has been thoroughly
investigated. Maher~\cite{Maher93alogic} presents fundamental LP
results translated to the context of CLP, and specifically formulates
many standard LP results in the context of CLP over finite trees. See
also Comon et al.~\cite{Comon_2001} for a survey of constraint solving
on terms, including finite and infinite trees. 

Unlike constraint-handling rules (CHR) based
approaches~\cite{Fruhwirth_2009}, our framework does not rewrite or
transform the constraint set when solving. Our approach does not
utilize rewrite rules; we take a \emph{semantic}, rather than
syntactic, approach~\cite{Comon1999}. Our solvers interpret
constraints directly in the term model to discover violations.

Schrijvers et al. offer a different motivation for separating
constraint solving and search~\cite{schrijvers2009monadic}. They
implement different advanced search strategies via monad transformers
over basic search monads. It's not yet clear where miniKanren's
interleaving DFS search fits into their framework, although this is a
topic we are currently investigating.

microKanren (and thus also miniKanren) is closely connected to pure
Prolog. Spivey and Seres~\cite{spivey1999embedding} embed a similar
subset of Prolog work on a Haskell embedding of Prolog. Kiselyov's
``Taste of Logic Programming''~\cite{Kiselyov2006taste}, and of course
Ralf Hinze's extensive work on implementations of Prolog-style
backtracking~\cite{hinze2000deriving,hinze2001prolog} are all closely
related.





There exists a different sort of CLP paradigm based on research in
constraint satisfaction problems using constraint propagation to
reduce the search space. cKanren, an earlier miniKanren for CLP, takes
this different approach and uses domain restriction and constraint
propagation~\cite{alvis2011minikanren}. Alvis et al. take as their
primary example finite domains. Unlike languages generated by our
framework, they minimize answer constraint sets and prettily format
the results.

\section{Conclusion}\label{conclusion}

We have presented a framework for developing microKanren-like CLP
languages in an instance of the CLP scheme. Decoupling the constraint
management from the inference, control, and variable management has
helped to clarify the behavior of microKanren. We support customary
miniKanren constraints as well as interesting and useful new ones.

In our implementation we deliberately reject certain common
optimizations and features that would have complicated our
implementation. The solvers we generate do not simplify the constraint
set as a consequence of solving. Indeed, these generated constraint
solvers are not at all specialized for incremental constraint solving,
to the point of even adding duplicate, wholly redundant,
constraints. We do not minimize even the answer constraint set, nor
project answers.

We do not intend to generate efficient, state-of-the-art CLP
languages, and on that front we have surely succeeded. Instead of
efficiency, our aim is a simple, general framework for implementing
constraints in microKanren. We envision our framework as a lightweight
tool for rapidly prototyping constraint sets. Language designers can
explore and test constraint definitions and interactions without
building or modifying a complicated and efficient dedicated solver. We
also imagine it as an educational artifact that provides functional
programmers a minimal executable instance of the CLP scheme.

Although we have preferred simplicity over performance here, we hope
to investigate the performance impacts of various simple optimizations
including incremental constraint solving, early
projection~\cite{fordan1999projection}, attributed
variables~\cite{le1990new}, or calling out to an appropriate dedicated
constraint solver. We hope to develop these optimizations as a series
of correctness-preserving transformations.



In future work we also hope to build an extensible, generic constraint
simplification framework analogous to our framework for building
constraint solvers. The language designer should have to write only
the individual constraint simplification rules for the framework to
produce a simplifier. Ideally this framework will infer an efficient
order in which to execute these simplification functions based on
abstract interpretation. We envision actually using a CHR approach in
these simplifying solvers. We also want to formalize the meaning of a
``kind'' of constraint-violation. Defining precisely what violations a
single violation predicate should check will clarify the language
designer's precise responsibilities.



As it exists, ours is a clear, simple framework for generating
miniKanren languages with constraints and serves as a test-bed for
developing constraint systems and an artifact of study. Further, it
serves as a foundation for continued future work in designing
constraint systems.

\section*{Acknowledgments}

We thank Will Byrd, Chung-chieh Shan, and Oleg Kiselyov for early
discussions of constraints in miniKanren. We thank Ryan Culpepper for
his improvements to the framework macros. We also thank our anonymous
reviewers for their suggestions and improvements.

\bibliographystyle{eptcs}
\bibliography{constraints}

\begin{thebibliography}{10}
\providecommand{\bibitemdeclare}[2]{}
\providecommand{\surnamestart}{}
\providecommand{\surnameend}{}
\providecommand{\urlprefix}{Available at }
\providecommand{\url}[1]{\texttt{#1}}
\providecommand{\href}[2]{\texttt{#2}}
\providecommand{\urlalt}[2]{\href{#1}{#2}}
\providecommand{\doi}[1]{doi:\urlalt{http://dx.doi.org/#1}{#1}}
\providecommand{\bibinfo}[2]{#2}

\bibitemdeclare{article}{alvis2011minikanren}
\bibitem{alvis2011minikanren}
\bibinfo{author}{Claire~E. \surnamestart Alvis\surnameend},
  \bibinfo{author}{Jeremiah~J. \surnamestart Willcock\surnameend},
  \bibinfo{author}{Kyle~M. \surnamestart Carter\surnameend},
  \bibinfo{author}{William~E. \surnamestart Byrd\surnameend} \&
  \bibinfo{author}{Daniel~P. \surnamestart Friedman\surnameend}
  (\bibinfo{year}{2011}): \emph{\bibinfo{title}{c{K}anren: mini{K}anren with
  Constraints}}.
\newblock {\sl \bibinfo{journal}{Scheme Workshop {\textquotesingle}11}}.

\bibitemdeclare{article}{brassel2008declaring}
\bibitem{brassel2008declaring}
\bibinfo{author}{Bernd \surnamestart Bra{\ss}el\surnameend},
  \bibinfo{author}{Sebastian \surnamestart Fischer\surnameend} \&
  \bibinfo{author}{Frank \surnamestart Huch\surnameend} (\bibinfo{year}{2008}):
  \emph{\bibinfo{title}{Declaring Numbers}}.
\newblock {\sl \bibinfo{journal}{Electronic Notes in Theoretical Computer
  Science}} \bibinfo{volume}{216}, pp. \bibinfo{pages}{111--124}.
\newblock \urlprefix\url{http://dx.doi.org/10.1016/j.entcs.2008.06.037}.

\bibitemdeclare{inproceedings}{alphamk}
\bibitem{alphamk}
\bibinfo{author}{William~E. \surnamestart Byrd\surnameend} \&
  \bibinfo{author}{Daniel~P. \surnamestart Friedman\surnameend}
  (\bibinfo{year}{2007}): \emph{\bibinfo{title}{\ak: A Fresh Name in Nominal
  Logic Programming}}.
\newblock In: {\sl \bibinfo{booktitle}{Proceedings of Scheme Workshop
  {\textquotesingle}07, Universit\'{e} Laval Technical Report DIUL-RT-0701}},
  pp. \bibinfo{pages}{79--90 (\textit{see also}
  \mbox{\url{http://webyrd.net/alphamk/alphamk.pdf}} \textit{for
  improvements})}.

\bibitemdeclare{inproceedings}{Byrd:2012:quines}
\bibitem{Byrd:2012:quines}
\bibinfo{author}{William~E. \surnamestart Byrd\surnameend},
  \bibinfo{author}{Eric \surnamestart Holk\surnameend} \&
  \bibinfo{author}{Daniel~P. \surnamestart Friedman\surnameend}
  (\bibinfo{year}{2012}): \emph{\bibinfo{title}{{miniKanren}, live and
  untagged}}.
\newblock In: {\sl \bibinfo{booktitle}{Proceedings of Scheme Workshop
  {\textquotesingle}12}}, \bibinfo{publisher}{ACM}.
\newblock \urlprefix\url{http://dx.doi.org/10.1145/2661103.2661105}.

\bibitemdeclare{incollection}{clark1978negation}
\bibitem{clark1978negation}
\bibinfo{author}{Keith~L. \surnamestart Clark\surnameend}
  (\bibinfo{year}{1978}): \emph{\bibinfo{title}{Negation as Failure}}.
\newblock In: {\sl \bibinfo{booktitle}{Logic and Data Bases}},
  \bibinfo{publisher}{Springer Science LNCS}, pp. \bibinfo{pages}{293--322}.
\newblock \urlprefix\url{http://dx.doi.org/10.1007/978-1-4684-3384-5\_11}.

\bibitemdeclare{incollection}{comon1994constraints}
\bibitem{comon1994constraints}
\bibinfo{author}{Hubert \surnamestart Comon\surnameend} (\bibinfo{year}{1994}):
  \emph{\bibinfo{title}{Constraints in Term Algebras (Short Survey)}}.
\newblock In: {\sl \bibinfo{booktitle}{Algebraic Methodology and Software
  Technology ({AMAST}'93)}}, \bibinfo{publisher}{Springer Science LNCS}, pp.
  \bibinfo{pages}{97--108}.
\newblock \urlprefix\url{http://dx.doi.org/10.1007/978-1-4471-3227-1\_9}.

\bibitemdeclare{article}{Comon1999}
\bibitem{Comon1999}
\bibinfo{author}{Hubert \surnamestart Comon\surnameend},
  \bibinfo{author}{Mehmet \surnamestart Dincbas\surnameend},
  \bibinfo{author}{Jean-Pierre \surnamestart Jouannaud\surnameend} \&
  \bibinfo{author}{Claude \surnamestart Kirchner\surnameend}
  (\bibinfo{year}{1999}): \emph{\bibinfo{title}{A Methodological View of
  Constraint Solving}}.
\newblock {\sl \bibinfo{journal}{Constraints}}
  \bibinfo{volume}{4}(\bibinfo{number}{4}), pp. \bibinfo{pages}{337--361}.
\newblock \urlprefix\url{http://dx.doi.org/10.1023/A:1009868906501}.

\bibitemdeclare{incollection}{Comon_2001}
\bibitem{Comon_2001}
\bibinfo{author}{Hubert \surnamestart Comon\surnameend} \&
  \bibinfo{author}{Claude \surnamestart Kirchner\surnameend}
  (\bibinfo{year}{2001}): \emph{\bibinfo{title}{Constraint Solving on Terms}}.
\newblock In: {\sl \bibinfo{booktitle}{Constraints in Computational Logics}},
  \bibinfo{publisher}{Springer Science LNCS}, pp. \bibinfo{pages}{47--103}.
\newblock \urlprefix\url{http://dx.doi.org/10.1007/3-540-45406-3\_2}.

\bibitemdeclare{article}{culpepper2010fortifying}
\bibitem{culpepper2010fortifying}
\bibinfo{author}{Ryan \surnamestart Culpepper\surnameend}
  (\bibinfo{year}{2012}): \emph{\bibinfo{title}{Fortifying macros}}.
\newblock {\sl \bibinfo{journal}{Journal of Functional Programming}}
  \bibinfo{volume}{22}(\bibinfo{number}{4-5}), pp. \bibinfo{pages}{439--476}.
\newblock \urlprefix\url{http://dx.doi.org/10.1017/s0956796812000275}.

\bibitemdeclare{techreport}{flatt2010reference}
\bibitem{flatt2010reference}
\bibinfo{author}{Matthew \surnamestart Flatt\surnameend} \&
  \bibinfo{author}{\surnamestart PLT\surnameend} (\bibinfo{year}{2010}):
  \emph{\bibinfo{title}{Reference: Racket}}.
\newblock \bibinfo{type}{Technical Report} \bibinfo{number}{PLT-TR-2010-1},
  \bibinfo{institution}{PLT Design Inc.}
\newblock \bibinfo{note}{\url{http://racket-lang.org/tr1/}}.

\bibitemdeclare{book}{fordan1999projection}
\bibitem{fordan1999projection}
\bibinfo{author}{Andreas \surnamestart Fordan\surnameend}
  (\bibinfo{year}{1999}): \emph{\bibinfo{title}{Projection in Constraint Logic
  Programming}}.
\newblock \bibinfo{publisher}{Ios Press}.

\bibitemdeclare{book}{Fruhwirth_2009}
\bibitem{Fruhwirth_2009}
\bibinfo{author}{Thom \surnamestart Fruhwirth\surnameend}
  (\bibinfo{year}{2009}): \emph{\bibinfo{title}{Constraint Handling Rules}}.
\newblock \bibinfo{publisher}{Cambridge University Press ({CUP})}.
\newblock \urlprefix\url{http://dx.doi.org/10.1017/cbo9780511609886}.

\bibitemdeclare{inproceedings}{hemann2016small}
\bibitem{hemann2016small}
\bibinfo{author}{Jason \surnamestart Hemann\surnameend},
  \bibinfo{author}{Daniel~P. \surnamestart Friedman\surnameend},
  \bibinfo{author}{William~E. \surnamestart Byrd\surnameend} \&
  \bibinfo{author}{Matthew \surnamestart Might\surnameend}
  (\bibinfo{year}{2016}): \emph{\bibinfo{title}{A small embedding of logic
  programming with a simple complete search}}.
\newblock In: {\sl \bibinfo{booktitle}{Proceedings of {DLS}
  {\textquotesingle}16}}, \bibinfo{publisher}{ACM}.
\newblock \urlprefix\url{http://dx.doi.org/10.1145/2989225.2989230}.

\bibitemdeclare{inproceedings}{hinze2000deriving}
\bibitem{hinze2000deriving}
\bibinfo{author}{Ralf \surnamestart Hinze\surnameend} (\bibinfo{year}{2000}):
  \emph{\bibinfo{title}{Deriving backtracking monad transformers}}.
\newblock In: {\sl \bibinfo{booktitle}{Proceedings of {ICFP}
  {\textquotesingle}00}}, \bibinfo{publisher}{ACM}.
\newblock \urlprefix\url{http://dx.doi.org/10.1145/351240.351258}.

\bibitemdeclare{article}{hinze2001prolog}
\bibitem{hinze2001prolog}
\bibinfo{author}{Ralf \surnamestart Hinze\surnameend} (\bibinfo{year}{2001}):
  \emph{\bibinfo{title}{Prolog's control constructs in a functional setting
  Axioms and implementation}}.
\newblock {\sl \bibinfo{journal}{International Journal of Foundations of
  Computer Science}} \bibinfo{volume}{12}(\bibinfo{number}{02}), pp.
  \bibinfo{pages}{125--170}, \doi{10.1142/S0129054101000436}.
\newblock \urlprefix\url{https://dx.doi.org/10.1142/S0129054101000436}.

\bibitemdeclare{inproceedings}{le1990new}
\bibitem{le1990new}
\bibinfo{author}{Serge~Le \surnamestart Huitouze\surnameend}
  (\bibinfo{year}{1990}): \emph{\bibinfo{title}{A new data structure for
  implementing extensions to Prolog}}.
\newblock In: {\sl \bibinfo{booktitle}{Programming Language Implementation and
  Logic Programming}}, \bibinfo{publisher}{Springer Science LNCS}, pp.
  \bibinfo{pages}{136--150}.
\newblock \urlprefix\url{http://dx.doi.org/10.1007/bfb0024181}.

\bibitemdeclare{inproceedings}{jaffar87constraint}
\bibitem{jaffar87constraint}
\bibinfo{author}{J.~\surnamestart Jaffar\surnameend} \& \bibinfo{author}{J.-L.
  \surnamestart Lassez\surnameend} (\bibinfo{year}{1987}):
  \emph{\bibinfo{title}{Constraint logic programming}}.
\newblock In: {\sl \bibinfo{booktitle}{Proceedings of {POPL}
  {\textquotesingle}87}}, \bibinfo{publisher}{ACM}.
\newblock \urlprefix\url{http://dx.doi.org/10.1145/41625.41635}.

\bibitemdeclare{article}{jaffar1998semantics}
\bibitem{jaffar1998semantics}
\bibinfo{author}{Joxan \surnamestart Jaffar\surnameend},
  \bibinfo{author}{Michael \surnamestart Maher\surnameend},
  \bibinfo{author}{Kim \surnamestart Marriott\surnameend} \&
  \bibinfo{author}{Peter \surnamestart Stuckey\surnameend}
  (\bibinfo{year}{1998}): \emph{\bibinfo{title}{The semantics of constraint
  logic programs}}.
\newblock {\sl \bibinfo{journal}{The Journal of Logic Programming}}
  \bibinfo{volume}{37}(\bibinfo{number}{1-3}), pp. \bibinfo{pages}{1--46}.
\newblock \urlprefix\url{http://dx.doi.org/10.1016/s0743-1066(98)10002-x}.

\bibitemdeclare{article}{jaffar1994constraint}
\bibitem{jaffar1994constraint}
\bibinfo{author}{Joxan \surnamestart Jaffar\surnameend} \&
  \bibinfo{author}{Michael~J. \surnamestart Maher\surnameend}
  (\bibinfo{year}{1994}): \emph{\bibinfo{title}{Constraint logic programming: a
  survey}}.
\newblock {\sl \bibinfo{journal}{The Journal of Logic Programming}}
  \bibinfo{volume}{19-20}, pp. \bibinfo{pages}{503--581}.
\newblock \urlprefix\url{http://dx.doi.org/10.1016/0743-1066(94)90033-7}.

\bibitemdeclare{article}{jaffar1993projecting}
\bibitem{jaffar1993projecting}
\bibinfo{author}{Joxan \surnamestart Jaffar\surnameend},
  \bibinfo{author}{Michael~J. \surnamestart Maher\surnameend},
  \bibinfo{author}{Peter~J. \surnamestart Stuckey\surnameend} \&
  \bibinfo{author}{Roland H.~C. \surnamestart Yap\surnameend}
  (\bibinfo{year}{1993}): \emph{\bibinfo{title}{Projecting \clpr constraints}}.
\newblock {\sl \bibinfo{journal}{New Gener Comput}}
  \bibinfo{volume}{11}(\bibinfo{number}{3-4}), pp. \bibinfo{pages}{449--469}.
\newblock \urlprefix\url{http://dx.doi.org/10.1007/bf03037187}.

\bibitemdeclare{inproceedings}{mkmacro}
\bibitem{mkmacro}
\bibinfo{author}{Andrew~W. \surnamestart Keep\surnameend},
  \bibinfo{author}{Michael~D. \surnamestart Adams\surnameend},
  \bibinfo{author}{Lindsey \surnamestart Kuper\surnameend},
  \bibinfo{author}{William~E. \surnamestart Byrd\surnameend} \&
  \bibinfo{author}{Daniel~P. \surnamestart Friedman\surnameend}
  (\bibinfo{year}{2009}): \emph{\bibinfo{title}{A Pattern-matcher for
  {miniKanren} -or- {How} to Get into Trouble with {CPS} Macros}}.
\newblock In: {\sl \bibinfo{booktitle}{Proceedings of Scheme Workshop
  {\textquotesingle}09, Cal Poly Technical Report CPSLO-CSC-09-03}}, pp.
  \bibinfo{pages}{37--45}.

\bibitemdeclare{misc}{Kiselyov2006taste}
\bibitem{Kiselyov2006taste}
\bibinfo{author}{Oleg \surnamestart Kiselyov\surnameend}
  (\bibinfo{year}{2006}): \emph{\bibinfo{title}{The taste of logic
  programming}}.
\newblock \urlprefix\url{http://okmij.org/ftp/Scheme/misc.html\#sokuza-kanren}.

\bibitemdeclare{inproceedings}{kiselyov2008pure}
\bibitem{kiselyov2008pure}
\bibinfo{author}{Oleg \surnamestart Kiselyov\surnameend},
  \bibinfo{author}{William~E. \surnamestart Byrd\surnameend},
  \bibinfo{author}{Daniel~P. \surnamestart Friedman\surnameend} \&
  \bibinfo{author}{{Chung-chieh} \surnamestart Shan\surnameend}:
  \emph{\bibinfo{title}{Pure, Declarative, and Constructive Arithmetic
  Relations (Declarative Pearl)}}.
\newblock In: {\sl \bibinfo{booktitle}{Functional and Logic Programming}},
  \bibinfo{publisher}{Springer Science LNCS}, pp. \bibinfo{pages}{64--80}.
\newblock \urlprefix\url{http://dx.doi.org/10.1007/978-3-540-78969-7\_7}.

\bibitemdeclare{inproceedings}{kiselyov2005backtracking}
\bibitem{kiselyov2005backtracking}
\bibinfo{author}{Oleg \surnamestart Kiselyov\surnameend},
  \bibinfo{author}{{Chung-chieh} \surnamestart Shan\surnameend},
  \bibinfo{author}{Daniel~P. \surnamestart Friedman\surnameend} \&
  \bibinfo{author}{Amr \surnamestart Sabry\surnameend} (\bibinfo{year}{2005}):
  \emph{\bibinfo{title}{Backtracking, interleaving, and terminating monad
  transformers: (functional pearl)}}.
\newblock In: {\sl \bibinfo{booktitle}{Proceedings of {ICFP}
  {\textquotesingle}05}}, \bibinfo{volume}{40}, \bibinfo{publisher}{ACM}, pp.
  \bibinfo{pages}{192--203}.
\newblock \urlprefix\url{http://doi.acm.org/10.1145/1086365.1086390}.

\bibitemdeclare{misc}{kumar2010mechanising}
\bibitem{kumar2010mechanising}
\bibinfo{author}{Ramana \surnamestart Kumar\surnameend} (\bibinfo{year}{2010}):
  \emph{\bibinfo{title}{Mechanising {A}spects of {miniKanren} in {HOL}}}.
\newblock \bibinfo{note}{Australian National University. Bachelors thesis}.

\bibitemdeclare{article}{lloyd1991partial}
\bibitem{lloyd1991partial}
\bibinfo{author}{J.~W. \surnamestart Lloyd\surnameend} \&
  \bibinfo{author}{J.~C. \surnamestart Shepherdson\surnameend}
  (\bibinfo{year}{1991}): \emph{\bibinfo{title}{Partial evaluation in logic
  programming}}.
\newblock {\sl \bibinfo{journal}{The Journal of Logic Programming}}
  \bibinfo{volume}{11}(\bibinfo{number}{3-4}), pp. \bibinfo{pages}{217--242}.
\newblock \urlprefix\url{http://dx.doi.org/10.1016/0743-1066(91)90027-m}.

\bibitemdeclare{inproceedings}{Maher93alogic}
\bibitem{Maher93alogic}
\bibinfo{author}{Michael~J. \surnamestart Maher\surnameend}
  (\bibinfo{year}{1993}): \emph{\bibinfo{title}{A Logic Programming View of
  {CLP}}}.
\newblock In: {\sl \bibinfo{booktitle}{Proceedings of {ICLP}
  {\textquotesingle}93}}, \bibinfo{publisher}{MIT Press}, pp.
  \bibinfo{pages}{737--753}.

\bibitemdeclare{inproceedings}{alphatap}
\bibitem{alphatap}
\bibinfo{author}{Joseph~P. \surnamestart Near\surnameend},
  \bibinfo{author}{William~E. \surnamestart Byrd\surnameend} \&
  \bibinfo{author}{Daniel~P. \surnamestart Friedman\surnameend}
  (\bibinfo{year}{2008}): \emph{\bibinfo{title}{\alphatap: A Declarative
  Theorem Prover for First-Order Classical Logic}}.
\newblock In: {\sl \bibinfo{booktitle}{Proceedings of {ICLP}
  {\textquotesingle}08}}, pp. \bibinfo{pages}{238--252}.
\newblock \urlprefix\url{http://dx.doi.org/10.1007/978-3-540-89982-2\_26}.

\bibitemdeclare{misc}{nolen2016core}
\bibitem{nolen2016core}
\bibinfo{author}{David \surnamestart Nolen\surnameend} (\bibinfo{year}{2016}):
  \emph{\bibinfo{title}{core.logic}}.
\newblock \bibinfo{howpublished}{\url{https://github.com/clojure/core.logic}}.

\bibitemdeclare{article}{schrijvers2009monadic}
\bibitem{schrijvers2009monadic}
\bibinfo{author}{Tom \surnamestart Schrijvers\surnameend},
  \bibinfo{author}{Peter \surnamestart Stuckey\surnameend} \&
  \bibinfo{author}{Philip \surnamestart Wadler\surnameend}
  (\bibinfo{year}{2009}): \emph{\bibinfo{title}{Monadic constraint
  programming}}.
\newblock {\sl \bibinfo{journal}{Journal of Functional Programming}}
  \bibinfo{volume}{19}(\bibinfo{number}{06}), p. \bibinfo{pages}{663}.
\newblock \urlprefix\url{http://dx.doi.org/10.1017/s0956796809990086}.

\bibitemdeclare{inproceedings}{spivey1999embedding}
\bibitem{spivey1999embedding}
\bibinfo{author}{J.~M. \surnamestart Spivey\surnameend} \&
  \bibinfo{author}{Silvija \surnamestart Seres\surnameend}
  (\bibinfo{year}{1999}): \emph{\bibinfo{title}{Embedding {P}rolog in
  {H}askell}}.
\newblock In \bibinfo{editor}{E.~\surnamestart Meier\surnameend}, editor: {\sl
  \bibinfo{booktitle}{Proceedings of Haskell Workshop {\textquotesingle}99,
  Utrecht University Technical Report UU-CS-1999-28}}.
\newblock
  \urlprefix\url{http://www.cs.uu.nl/research/techreps/repo/CS-1999/1999-28.pdf}.

\end{thebibliography}

\newpage

\appendix

\section*{Appendix: unification and microKanren control}


Below are the implementation of \verb|unify| and microKanren's control
and variable management infrastructure. The first five lines implement
variables and variable management. The next twenty lines implement
substitutions and \verb|unify|. The following six lines are the macro
that implements \verb|define-relation|, which plays a part in the
interleaving, and the operators for conjunction and disjunction. The
next twelve lines implement help functions for \verb|conj| and
\verb|disj| that interleave streams. The subsequent twelve lines
define \verb|ifte| and \verb|once|, impure microKanren operators that
provide soft-cut and committed choice, respectively. The final ten
lines actually run the computation, forcing a stream to mature, and
also enable us to get a prefix from the mature, computed stream.

\begin{multicols}{2}  
{\footnotesize
\begin{Verbatim}[commandchars=\\\{\},codes={\catcode`$=3\catcode`^=7\catcode`_=8},formatcom={\lstset{fancyvrb=false}}]
(define (var n) n)

(define (var? n) (number? n))

(define ((call/fresh f) S/c)
  (let ((S (car S/c)) (c (cdr S/c)))
    ((f (var c)) `(,S . ,(+ 1 c)))))

(define (occurs? x v s)
  (let ((v (walk v s)))
    (cond ((var? v) (eqv? x v))
          ((pair? v) (or (occurs? x (car v) s)
                         (occurs? x (cdr v) s)))
          (else #f))))

(define (ext-s x v s) 
  (if (occurs? x v s) #f `((,x . ,v) . ,s)))

(define (walk u s)
  (let ((pr (assv u s)))
    (if pr (walk (cdr pr) s) u)))

(define (unify u v s)
  (let ((u (walk u s)) (v (walk v s)))
    (cond ((eqv? u v) s)
          ((var? u) (ext-s u v s))
          ((var? v) (ext-s v u s))
          ((and (pair? u) (pair? v))
           (let ((s (unify (car u) (car v) s)))
             (and s (unify (cdr u) (cdr v) s))))
          (else #f))))

(define-syntax-rule (define-relation (r . as) g)
  (define ((r . as) S/c) (delay/name (g S/c))))

(define ((disj g\(\mathtt{_1}\) g\(\mathtt{_2}\)) S/c) 
  (\$append (g\(\mathtt{_1}\) S/c) (g\(\mathtt{_2}\) S/c)))

(define ((conj g\(\mathtt{_1}\) g\(\mathtt{_2}\)) S/c) 
  (\$append-map g\(\mathtt{_2}\) (g\(\mathtt{_1}\) S/c)))

(define (\$append \$\(\mathtt{_1}\) \$\(\mathtt{_2}\))
  (cond ((null? \$\(\mathtt{_1}\)) \$\(\mathtt{_2}\))
        ((promise? \$\(\mathtt{_1}\)) 
         (delay/name (\$append \$\(\mathtt{_2}\) (force \$\(\mathtt{_1}\)))))
        (else (cons (car \$\(\mathtt{_1}\)) 
                (\$append (cdr \$\(\mathtt{_1}\)) \$\(\mathtt{_2}\))))))

(define (\$append-map g \$)
  (cond ((null? \$) `())
        ((promise? \$) 
         (delay/name (\$append-map g (force \$))))
        (else (\$append (g (car \$)) 
               (\$append-map g (cdr \$))))))
  
(define ((ifte g\(\mathtt{_1}\) g\(\mathtt{_2}\) g\(\mathtt{_3}\)) s/c)
  (let loop ((\$ (g\(\mathtt{_1}\) s/c)))
    (cond
      ((null? \$) (g\(\mathtt{_3}\) s/c))
      ((promise? \$) (delay/name (loop (force \$))))
      (else (\$append-map \$ g\(\mathtt{_2}\))))))

(define ((once g) s/c)
  (let loop ((\$ (g s/c)))
    (cond
      ((null? \$) '())
      ((promise? \$) (delay/name (loop (force \$))))
      (else (list (car \$))))))

(define (pull \$) 
  (if (promise? \$) (pull (force \$)) \$))

(define (take n \$)
  (cond ((null? \$) '())
        ((and n (zero? (- n 1))) (list (car \$)))
        (else (cons (car \$) 
                (take (and n (- n 1)) 
                      (pull (cdr \$)))))))

(define (call/initial-state n g)
  (take n (pull (g `(,S\(\mathtt{_0}\) . 0)))))

\end{Verbatim}
}
\end{multicols}

\end{document}